\documentstyle[preprint,aps]{revtex}

\begin{document}
\draft
 
\title{Novel soft-photon analysis of pp\mbox{\boldmath$\!\!\gamma$}
       below \\ pion-production threshold}

\author{M.K. Liou}
\address{Department of Physics and Institute for Nuclear Theory, \\
         Brooklyn College of CUNY, Brooklyn, NY 11210} 
\author{R. Timmermans and B.F. Gibson}
\address{Theoretical Division, Los Alamos National Laboratory,
         Los Alamos, NM 87545}

\date{\today}

\maketitle

\begin{abstract}
A novel soft-photon amplitude is proposed to replace the
conventional Low soft-photon amplitude for nucleon-nucleon
bremsstrahlung. Its derivation is guided by the standard
meson-exchange model of the nucleon-nucleon interaction.
This new amplitude provides a superior description of
$pp\gamma$ data. The predictions of this new amplitude
are in close agreement with potential-model calculations,
which implies that, contrary to conclusions drawn by
others, off-shell effects are essentially insignificant
below pion-production threshold.
\end{abstract}

Bremsstrahlung processes have been used as a tool to
investigate electromagnetic properties of resonances,
details of reaction mechanisms, and off-shell properties
of scattering amplitudes. The most succesful example
in the first case is the determination of the magnetic
moments of the $\Delta^{++}$ ($\Delta^0$) from $\pi^+p\gamma$
($\pi^-p\gamma$) data in the energy region of the $\Delta$(1232)
resonance~\cite{Lin91}. In the case of reaction mechanisms,
a well-known example is the extraction of nuclear time
delays from the $p ^{12}C\gamma$ data near the 1.7-MeV
resonance~\cite{Mar76}. The time delay distinguishes
between direct and compound nuclear reactions. The initial
goal of nucleon-nucleon bremsstrahlung investigations was
to distinguish among various phenomenological potential
models of the fundamental two-nucleon interaction. Most
measured $pp\gamma$ cross sections could, in fact, be reasonably
described by potential-model calculations, but the difference
between predictions from any two realistic potentials
appears to be too small to be distinguished by the data.

For more than 30 years, the conventional Low soft-photon
amplitude~\cite{Low58} has been widely used for studying
nuclear and particle bremsstrahlung processes. It seemingly
provides a good description of the data for some processes.
For instance, Nyman~\cite{Nym68} and Fearing~\cite{Fea80}
used this amplitude to calculate $pp\gamma$ cross
sections which were in reasonable agreement with
several measurements and potential-model calculations.
However, it was recently pointed out by Workman and
Fearing~\cite{Wor86} that the results from this conventional
Low amplitude differ significantly from the potential-model
calculations for the TRIUMF data at 280 MeV~\cite{Mic90}. This
difference was interpreted as evidence for ``off-shell effects''
in the $pp\gamma$ process.

The main purpose of this Letter is to propose a novel
soft-photon amplitude to replace the conventional Low
prescription. This new amplitude, the derivation of which
is guided by the structure of the standard meson-exchange model
of the two-nucleon interaction, is relativistic, manifestly
gauge invariant, and consistent with the soft-photon
theorem. It belongs to one of the two general classes
of recently derived soft-photon amplitudes~\cite{Lio93}.
We demonstrate that the $pp\gamma$ data from low energies
to energies near the pion-production threshold can be
consistently described by the new amplitude. Most
importantly, we point out here that our amplitude
essentially eliminates the discrepancy between the
soft-photon approximation and the potential-model
calculations. That is, we demonstrate that
``off-shell effects'' are essentially
negligible. Finally, we explore why the conventional
Low amplitude works for some cases but fails for others.

In order to elucidate these points, let us consider
photon emission accompanying the scattering of two
spin-1/2 particles $A$ and $B$,
\begin{equation}
   A(q_i^\mu) + B(p_i^\mu) \rightarrow
   A(q_f^\mu) + B(p_f^\mu) + \gamma(K^\mu) \ . \label{eq:ppg}
\end{equation}
Here, $q_i^\mu$ ($q_f^\mu$) and $p_i^\mu$ ($p_f^\mu$)
are the initial (final) four-momenta for particles
$A$ and $B$, respectively, and $K^\mu$ is the four-momentum
for the emitted photon with polarization $\varepsilon^\mu$.
Particle $A$ ($B$) is assumed to have mass $m_A$ ($m_B$),
charge $Q_A$ ($Q_B$), and anomalous magnetic moment
$\kappa_A$ ($\kappa_B$). For process~(\ref{eq:ppg}),
we can define the following Mandelstam variables:
$s_i=(q_i+p_i)^2$, $s_f=(q_f+p_f)^2$, $t_q=(q_f-q_i)^2$,
$t_p=(p_f-p_i)^2$, $u_1=(p_f-q_i)^2$, and $u_2=(q_f-p_i)^2$.
Since a soft-photon amplitude depends only on either
($s$,$t$) or ($u$,$t$), chosen from the above set,
we can derive two distinct classes of
soft-photon amplitudes: $M^{(1)}_\mu(s,t)$ and
$M^{(2)}_\mu(u,t)$~\cite{Lio93}. The general amplitude from
the first class is the two-$s$--two-$t$ special amplitude
$M^{TsTts}_\mu(s_i,s_f;t_q,t_p)$; that from the second
class is the two-$u$--two-$t$ special amplitude
$M^{TuTts}_\mu(u_1,u_2;t_q,t_p)$. The distinguishing
characteristics of these amplitudes come from the fact
that they are evaluated at different elastic-scattering
or on-shell points (energy and angle). The soft-photon theorem
does not specify how these on-shell points are to be selected.

The modified procedure
for deriving these soft-photon amplitudes is described
in detail in Ref.~\cite{Lio93}. In this procedure, the
fundamental tree diagrams of the underlying elastic
scattering process play an important role in deriving
the two general amplitudes. Thus, we argue that $M^{TsTts}_\mu$
should be used to describe those processes which are
resonance dominated [such as $p ^{12}C\gamma$ near 1.7 MeV
and $\pi^\pm p\gamma$ in the $\Delta$(1232) region], whereas
$M^{TuTts}_\mu$ should be used to describe those processes
which are exchange-current dominated (such as the $np\gamma$
process). For the $pp\gamma$ process, which exhibits
neither strong resonance effects nor significant $u$-channel
exchange-current effects, both amplitudes can be used in
theory, although this has never been tested in conjunction
with experimental data. We provide here the results of such
an analysis. We emphasize that the general amplitude
$M^{TuTts}_\mu$ (not $M^{TsTts}_\mu$) arises naturally
for nucleon-nucleon bremsstrahlung {\it if} the derivation is
guided by the standard meson-exchange model of the two-nucleon
interaction.

The amplitude $M^{TuTts}_\mu$ for the $pp\gamma$
process can be written in terms of five invariant amplitudes
$F^e_\alpha$ ($\alpha=1,\ldots,5$) as
\begin{equation}
   M^{TuTts}_\mu = \sum_{\alpha=1}^5 \left[
  Q_A\overline{u}(q_f)X_{\alpha\mu}u(q_i)\overline{u}(p_f)g^\alpha u(p_i)
  + Q_B\overline{u}(q_f)g_\alpha u(q_i)\overline{u}(p_f)Y^\alpha_\mu u(p_i)
  \right] \ , \label{eq:MTuTt}
\end{equation}
where
\begin{eqnarray}
  X_{\alpha\mu} &=& F^e_\alpha(u_1,t_p)\left[
\frac{q_{f\mu}+R^{q_f}_\mu}{q_f\cdot K}-\frac{(p_i-q_f)_\mu}{(p_i-q_f)\cdot K}
  \right] g_\alpha \nonumber \\ && - F^e_\alpha(u_2,t_p) g_\alpha \left[
\frac{q_{i\mu}+R^{q_i}_\mu}{q_i\cdot K}-\frac{(q_i-p_f)_\mu}{(q_i-p_f)\cdot K}
\right] \ , \label{eq:Xamu} \\
  Y^\alpha_\mu &=& F^e_\alpha(u_2,t_q)\left[
\frac{p_{f\mu}+R^{p_f}_\mu}{p_f\cdot K}-\frac{(q_i-p_f)_\mu}{(q_i-p_f)\cdot K}
  \right] g^\alpha  \nonumber \\ &&- F^e_\alpha(u_1,t_q) g^\alpha \left[
\frac{p_{i\mu}+R^{p_i}_\mu}{p_i\cdot K}-\frac{(p_i-q_f)_\mu}{(p_i-q_f)\cdot K}
\right] \  . \label{eq:Yamu}
\end{eqnarray}
In Eqs.~(\ref{eq:MTuTt}-\ref{eq:Yamu}), we have defined
\begin{eqnarray}
 (g_1,g_2,g_3,g_4,g_5) & \equiv &
 (1,\sigma_{\mu\nu}/\sqrt{2},i\gamma_5\gamma_\mu,\gamma_\mu,\gamma_5)
\ , \nonumber \\
 (g^1,g^2,g^3,g^4,g^5) & \equiv &
 (1,\sigma^{\mu\nu}/\sqrt{2},i\gamma_5\gamma^\mu,\gamma^\mu,\gamma_5)
\ , \nonumber
\end{eqnarray}
and the factors $R^Q_\mu$ ($Q=q_f,q_i,p_f,p_i$) can be expressed as
\begin{equation}
  R^Q_\mu = \frac{1}{4}\left[\gamma_\mu,\slash\hspace{-.28cm}K\right] +
\frac{\kappa}{8m}\left\{\left[\gamma_\mu,\slash\hspace{-.28cm}K\right],
   \slash\hspace{-.25cm}Q\right\} \ .
\label{eq:RQmu}
\end{equation}
In Eq.~(\ref{eq:RQmu}), $m$ ($=m_A=m_B$) and $\kappa$ ($=\kappa_A=\kappa_B$)
are the mass and the anomalous magnetic moment of the proton,
$\slash\hspace{-.25cm}Q=Q^\mu\gamma_\mu$, and we have used
$\left[F,G\right] \equiv FG-GF$ and $\left\{F,G\right\} \equiv FG+GF$.
As one can see from Eqs.~(\ref{eq:Xamu}) and (\ref{eq:Yamu}), the
invariant amplitudes $F^e_\alpha$ depend on $u$ and $t$. The same
amplitudes but as functions of $s$ and $t$ can be obtained if
we use the condition $s+t+u=4m^2$. For example,
$F^e_\alpha(u_1,t_p)=F^e_\alpha(s_{1p},t_p)$ where $s_{1p}+t_p+u_1=4m^2$.
Since $F^e_\alpha(s_{1p},t_p)$ ($\alpha=1,\ldots,5$) are invariant
amplitudes for the $pp$ elastic process, the Feynman amplitude
$F(s_{1p},t_p)$ defined by Goldberger {\it et al}.~\cite{Gol60} can
be written in terms of the five Fermi covariants ($S,T,A,V,P$) as
\begin{equation}
 F(s_{1p},t_p) =
       F^e_1(s_{1p},t_p)S+F^e_2(s_{1p},t_p)T+F^e_3(s_{1p},t_p)A
             +F^e_4(s_{1p},t_p)V+F^e_5(s_{1p},t_p)P \ .
\end{equation}

The amplitude $M^{TsTts}_\mu(s_i,s_f;t_q,t_p)$ can be formally
obtained from the amplitude $M^{TuTts}_\mu(u_1,u_2;t_q,t_p)$ given
by Eqs.~(\ref{eq:MTuTt}), (\ref{eq:Xamu}), and (\ref{eq:Yamu})
by making the following substitutions: ($i$) $Q_B\rightarrow -Q_B$
and ($ii$) $p_i^\mu \leftrightarrow -p_f^\mu$ and $g^\alpha R^{p_i}_\mu
\leftrightarrow -R^{p_f}_\mu g^\alpha$,
keeping $R^{q_i}_\mu$, $R^{q_f}_\mu$,
and the spinors $\overline{u}$ and $u$ unchanged. However,
we emphasize that the two are not the same numerically.

If all $F^e_\alpha(s_x,t_y)$ ($\alpha=1,\ldots,5$, $x=i,f$, and
$y=q,p$) in $M^{TsTts}_\mu$ are expanded about average $s$,
$\overline{s}$, and average $t$, $\overline{t}$, then the first
two terms of the expansion give the conventional Low amplitude
$M^{{\rm Low}(s,t)}_\mu(\overline{s},\overline{t})$. This
particular choice ($\overline{s},\overline{t}$) for the
on-shell point at which the Low amplitude is evaluated
is just an {\it ad hoc} prescription, although it provided
a reasonable description of $pp\gamma$ data until the
TRIUMF measurements at 280 MeV.

We have studied the amplitudes $M^{TuTts}_\mu$,
$M^{TsTts}_\mu$, and $M^{{\rm Low}(s,t)}_\mu$ and have
applied them to calculate $pp\gamma$ cross sections at various
energies, using state-of-the-art phase shifts from the latest
Nijmegen $pp$ partial-wave analysis~\cite{Ber90}. Anecdotal
results are shown in Figs.~\ref{fig1}, \ref{fig2},
and \ref{fig3}. At 42 MeV for
$\theta_q=\theta_p=26^\circ$ (see Fig.~\ref{fig1}) the
coplanar cross sections calculated from $M^{TsTts}_\mu$ are
much larger than the Manitoba data~\cite{Jov71}. The amplitudes
$M^{TuTts}_\mu$ and $M^{{\rm Low}(s,t)}_\mu$, on the other
hand, give similar results which agree well with both
the data (within the experimental error)
and the representative Hamada-Johnston-potential
calculation~\cite{Lio72}. The results calculated
using $M^{{\rm Low}(s,t)}_\mu$ are close to those
obtained by Nyman and Fearing. In Fig.~\ref{fig2}
our coplanar cross sections calculated from
$M^{TuTts}_\mu$ and $M^{{\rm Low}(s,t)}_\mu$ at
157 MeV for $\theta_q=\theta_p=35^\circ$ are
compared with the Harvard data~\cite{Got67} and
a Paris-potential calculation~\cite{Jet93}. (Other
potential-model calculations~\cite{Wor86,Her92,Bro92,Kat93}
which include relativistic spin-corrections etc.\
are similar.) Cross sections calculated
using the amplitude $M^{TsTts}_\mu$ are missing
from Figs.~\ref{fig2} and \ref{fig3}, because they
are factors larger than those plotted. Again
the amplitudes $M^{TuTts}_\mu$ and $M^{{\rm Low}(s,t)}_\mu$
give very similar results at this energy and agree
reasonably with both the potential-model curve
and the Harvard data.

However,
at an energy near the pion-production threshold and far
from the on-shell point, the two amplitudes $M^{TuTts}_\mu$
and $M^{{\rm Low}(s,t)}_\mu$ predict quite different
results. This is demonstrated in Fig.~\ref{fig3}. At 280 MeV
for $\theta_q=12.4^\circ$ and $\theta_p=12^\circ$, the
curve calculated from $M^{TuTts}_\mu$ agrees well with
the published TRIUMF data~\cite{Mic90} and with the
curves calculated using the Paris potential and the
Bonn potential~\cite{Mic90}. The amplitude
$M^{{\rm Low}(s,t)}_\mu$, on the other hand, predicts cross
sections which are too small for forward ($\theta_\gamma
\leq 30^\circ$) and backward ($\theta_\gamma \geq 150^\circ$)
photon angles. That $M^{{\rm Low}(s,t)}_\mu$ can describe
most of the older $pp\gamma$ data but fails to fit the new
TRIUMF data has already been pointed out by Fearing.
What is emphasized here is that the new amplitude
$M^{TuTts}_\mu$ describes data where the conventional Low
amplitude $M^{{\rm Low}(s,t)}_\mu$ fails. In other words,
the correct soft-photon amplitude which describes
the $pp\gamma$ data consistently is $M^{TuTts}_\mu$.

How can we understand the failure of the conventional
Low amplitude $M^{{\rm Low}(s,t)}_\mu$? Consider the
expressions given in Eqs.~(\ref{eq:MTuTt}-\ref{eq:Yamu}).
If we impose the on-shell condition, $s+t+u=4m^2$,
we can write $F^e_\alpha(u_1,t_p)=F^e_\alpha(s_{1p},t_p)$,
$F^e_\alpha(u_2,t_p)=F^e_\alpha(s_{2p},t_p)$,
$F^e_\alpha(u_1,t_q)=F^e_\alpha(s_{1q},t_q)$, and
$F^e_\alpha(u_2,t_q)=F^e_\alpha(s_{2q},t_q)$, where
$s_{1p}=s_i-2q_f\cdot K$, $s_{2p}=s_i-2p_i\cdot K$,
$s_{2q}=s_i-2p_f\cdot K$, and $s_{1q}=s_i-2q_i\cdot K$.
This shows that $F^e_\alpha$ will be evaluated at
four different energies and four different angles
in constructing $M^{TuTts}_\mu$. (Potential-model
calculations also use four-energy-four-angle amplitudes.)
In contrast, $M^{TsTts}_\mu$ is evaluated at two
energies and four angles, while $M^{{\rm Low}(s,t)}_\mu$
is evaluated at just one energy and one angle.
To be specific, at 100 MeV
for $\theta_q=\theta_p=\theta_\gamma=30^\circ$,
we have $s_{1p}=3.648$ GeV$^2$, $s_{2q}=3.640$ GeV$^2$,
$s_{2p}=3.632$ GeV$^2$, and $s_{1q}=3.655$ GeV$^2$,
whereas $s_i=3.709$ GeV$^2$ and $s_f=3.578$ GeV$^2$, and
finally $\overline{s}=3.644$ GeV$^2$. These quantities are
the dominant factors determining the calculated
cross sections. Since $s_{1p} \simeq s_{2p} \simeq
s_{2q} \simeq s_{1q} \simeq \overline{s}$ (the
differences in c.m. energy between $s_{1p}$, $s_{2p}$,
$s_{2q}$, and $s_{1q}$ on one hand, and $\overline{s}$
on the other hand, are less than
about 3 MeV), $M^{{\rm Low}(s,t)}_\mu$ and
$M^{TuTts}_\mu$ predict similar results at
energies lower than 100 MeV and for large proton
angles. However, the value of $s_i$ is much larger
than the value of $s_f$. This is equivalent to a c.m.
energy difference of some 34 MeV. This large difference between
$s_i$ and $s_f$ is the primary reason for the huge cross sections
predicted by $M^{TsTts}_\mu$. As the incident energy increases
(or the proton angles decrease), the values of the four energies,
$s_{1p}$, $s_{2p}$, $s_{1q}$, and $s_{2q}$, will no longer
be close to one another, and they differ significantly from
$\overline{s}$, as well as $s_i$ and $s_f$. Thus, the cross
sections calculated using the amplitude $M^{TuTts}_\mu$
will differ from those calculated using either
$M^{{\rm Low}(s,t)}_\mu$ or $M^{TsTts}_\mu$. A more
systematic analysis, including other relevant factors,
will be given elsewhere.

In conclusion, we have demonstrated that
the amplitude $M^{TuTts}_\mu$, not the
conventional Low amplitude $M^{{\rm Low}(s,t)}_\mu$
nor the amplitude $M^{TsTts}_\mu$, is the correct
soft-photon amplitude to be used in describing
the nucleon-nucleon bremsstrahlung processes.
Furthermore, below the pion-production threshold this
novel amplitude $M^{TuTts}_\mu$ provides a description
of the $pp\gamma$ data that is the equal of
contemporary potential-model calculations,
implying off-shell effects are insignificant
in the kinematic range measured to date.

\acknowledgments
We would like to thank M. Rentmeester for providing
the phase shifts from the latest Nijmegen proton-proton
partial-wave analysis. The work of M.K.L. was supported
in part by the City University of New York Professional
Staff Congress-Board of Higher Education Faculty Research
Award Program, while the work of R.T. and B.F.G. was
performed under the auspices of the Department of Energy.

\tighten
\newpage
\begin{figure}
\caption{Coplanar $pp\gamma$ cross section at 42 MeV for
         $\theta_q=\theta_p=26^\circ$; ------: result
         using $M^{TuTts}_\mu$; -- $\cdot$ -- $\cdot$:
         result using $M^{{\rm Low}(s,t)}_\mu$; -- -- --:
         result using $M^{TsTts}_\mu$; $\cdots\cdots$: result
         for Hamada-Johnston potential~\protect\cite{Lio72}.
         The data are from Ref.~\protect\cite{Jov71}.}
\label{fig1}
\end{figure}

\begin{figure}
\caption{Coplanar $pp\gamma$ cross section at 157 MeV for
         $\theta_q=\theta_p=35^\circ$; ------: result using
         $M^{TuTts}_\mu$; -- $\cdot$ -- $\cdot$:
         result using $M^{{\rm Low}(s,t)}_\mu$;
         -- -- --: result for Paris potential~\protect\cite{Jet93}.
         The data are from Ref.~\protect\cite{Got67}.}
\label{fig2}
\end{figure}

\begin{figure}
\caption{Coplanar $pp\gamma$ cross section at 280 MeV for
         $\theta_q=12.4^\circ, \theta_p=12^\circ$; ------:
         result using $M^{TuTts}_\mu$; -- $\cdot$ -- $\cdot$:
         result using $M^{{\rm Low}(s,t)}_\mu$; -- -- --:
         result for Paris potential~\protect\cite{Mic90};
         $\cdots\cdots$: result for Bonn potential~\protect\cite{Mic90}.
         The data are from Ref.~\protect\cite{Mic90}.}
\label{fig3}
\end{figure}

\end{document}